\documentclass[epj]{webofc}
\usepackage[varg]{txfonts}   
\usepackage{bm} 
\woctitle{Physics Opportunities at an Electron-Ion Collider}
\begin{document}
\title{Electron--deuteron DIS with spectator tagging at EIC: \\
Development of theoretical framework}
%
%
%
\author{
W.~Cosyn\inst{1} \and
V.~Guzey\inst{2}  \and
M.~Sargsian\inst{3} \and
M.~Strikman\inst{4} \and
C.~Weiss \inst{5}
}
\institute{
Ghent University, 9000 Ghent, Belgium \and
Petersburg Nuclear Physics Institute, Gatchina, 188300, Russia \and
Florida International University, Miami, FL 33199, USA \and
Pennsylvania State University, University Park, PA 16802, USA \and
Jefferson Lab, Newport News, VA 23606, USA
}
\abstract{An Electron--Ion Collider (EIC) would enable next-generation measurements
of deep-inelastic scattering (DIS) on the deuteron with detection of a forward-moving 
nucleon ($p, n$) and measurement of its recoil momentum (``spectator tagging''). 
Such experiments offer full control of the nuclear configuration during the high-energy 
process and can be used for precision studies of the neutron's partonic structure and
its spin dependence,
nuclear modifications of partonic structure, and nuclear shadowing at small $x$. 
We review the theoretical description of spectator tagging at EIC energies 
(light--front nuclear structure, on-shell extrapolation in the recoil nucleon momentum, 
final-state interactions, diffractive effects at small $x$) and report about 
on-going developments.}
\maketitle
Measurements of deep-inelastic scattering and related processes on light ions ($A = 2, 3, ...$)
address basic questions of high-energy nuclear physics and Quantum Chromodynamics (QCD), 
such as the partonic structure of the neutron and its spin decomposition \cite{Aidala:2012mv}, 
the nuclear modification of the nucleon's quark and gluon densities 
\cite{Frankfurt:1988nt,Malace:2014uea}, and the onset of nuclear shadowing 
and coherent phenomena at small $x$ \cite{Frankfurt:2011cs}. 
The common challenge in the analysis of such measurements
is to account for the multitude of nuclear configurations 
present during the high-energy scattering process. 
The scattering can involve any of the constituent nucleons ($p, n$) in different states of 
their quantum--mechanical motion (momentum, spin), as well as non-nucleonic degrees of 
freedom excited by the nuclear binding. In the extraction of neutron structure one
needs to isolate the cross section arising from scattering on the neutrons and eliminate
the effects of nuclear binding (Fermi motion, non-nucleonic degrees of freedom). 
For neutron spin structure one must also infer the effective polarization of the
neutron and account for the polarization of non-nucleonic degrees of freedom, 
particularly intrinsic $\Delta$ isobars in polarized ${}^3$He \cite{Frankfurt:1996nf}. 
In the study of nuclear modifications at $x \gtrsim 0.1$ (EMC effect, antishadowing) 
one would like to relate the modifications to the interactions taking place in a particular 
configuration (exchange mechanisms, short-range correlations). In the analysis of coherent
phenomena at small $x$ one wants to understand what configurations in the wave function
build up the coherent response (onset of shadowing, saturation). While these challenges
can partly be addressed by theoretical calculations, major progress could come from
new experiments that provide information on the nuclear configurations during the
high--energy process.

{\bf Deuteron and spectator tagging.} 
High-energy scattering on the deuteron ($A = 2$) with detection of a proton or neutron
in the nuclear fragmentation region (recoil momentum $p_R \sim$ few 10--100 MeV/$c$ in the 
deuteron rest frame) represents a unique method for performing DIS measurements in controlled 
nuclear configurations. The deuteron's $pn$ wave function is simple and well known 
from low--energy measurements, including the light--front
wave function entering in high--energy processes (see below). Because the deuteron has isospin 
$I = 0$, $\Delta$ isobars in the wave function are strongly suppressed (they can occur only 
in $\Delta\Delta$ configurations), so that the deuteron can be treated as a $pn$ system for 
most of the configurations relevant to DIS \cite{Frankfurt:1981mk}. Detection of the 
``spectator'' nucleon and measurement of its momentum positively identifies the
active nucleon and controls its momentum during the DIS process. Because of the
simple geometry of high-energy scattering on the $A = 2$ system the possibilities 
for final--state interactions are limited, and in configurations where they occur they can 
be computed using theoretical models. Regarding polarization the deuteron has spin $S = 1$; 
its wave function is predominantly S--wave with a small admixture of D--wave, so that the
effective nucleon polarization is well known.

DIS on the deuteron with recoil proton detection (``spectator tagging'') was measured
in the JLab CLAS BONuS experiment at 6 GeV beam energy at recoil momenta 
$p_R \gtrsim$ 70 MeV/$c$ \cite{Tkachenko:2014byy} and will be explored further 
at 11 GeV. In such fixed--target experiments it is difficult to get the slow proton 
(or neutron) out of the target and measure its momentum with sufficient resolution.
Much more suitable for this purpose are colliding--beam experiments, where the
spectator nucleon moves on with approximately half the deuteron beam momentum and
can be detected using forward detectors. An EIC with electron--nucleon center--of--mass 
energies $\surd s_{eN} \sim$ 15--100 GeV and luminosity 
$L \sim 10^{33}$--$10^{34}\, \textrm{cm}^{-2} \, \textrm{s}^{-1}$ is projected
as a future facility for nuclear physics \cite{2015NSAC}. Both EIC designs presently 
developed have capabilities for forward nucleon detection \cite{EIC-designs}. 
The JLab EIC detector is designed to 
provide full coverage for spectator protons down to zero transverse momentum,
and with a momentum resolution corresponding to $p_R \sim$ 20 MeV/$c$ in the
rest frame, as well as forward neutron detection \cite{detector}. 
This setup would enable detailed 
measurements of DIS with spectator tagging over the entire $(x, Q^2)$ range covered
by the collider, which includes the region of sea quarks, gluons, and small--$x$ phenomena.
It would allow also tagged measurements on the polarized deuteron, which is potentially
the most precise method for determining neutron spin structure. Altogether
this setup would permit nuclear DIS measurements with full control of the nuclear
configuration and enable a new level of understanding of nuclear effects in QCD.

To realize this potential it is necessary to develop the theoretical apparatus for 
describing DIS with spectator tagging at collider energies and with polarized beams. 
In this article we summarize the theoretical description and report about on-going 
developments \cite{LD1506}. A general review of nuclear physics
with EIC is given in Ref.~\cite{Accardi:2011mz}.

\begin{figure}
\centering
\includegraphics[width=.9\textwidth]{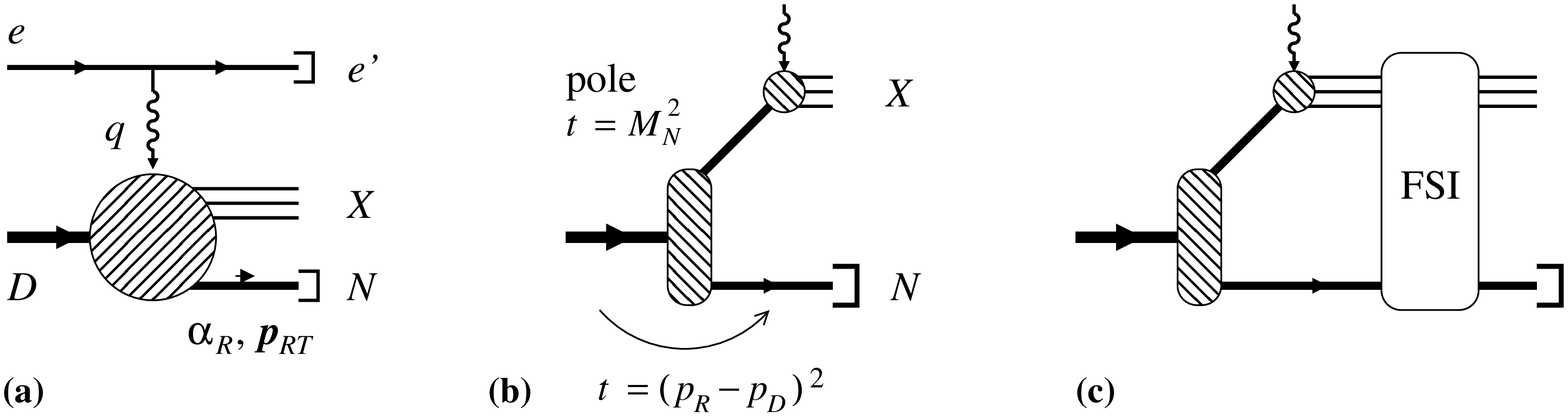}
\caption{(a) Tagged DIS on the deuteron, $e D \rightarrow e' + N + X$.
(b) Impulse approximation amplitude, containing the nucleon pole at $t = M_N^2$.
(c) Final--state interaction amplitude.}
\label{fig:tagged}
\end{figure}

{\bf Cross section and structure functions.} 
The basic observable is the invariant differential cross section for inclusive DIS 
on the deuteron with an identified nucleon with recoil momentum $p_R$ in the final state
(see Fig.~\ref{fig:tagged}a):
\begin{equation}
\frac{d\sigma (eD \rightarrow e' N X)}{dx \, dQ^2 \, (d^3 p_R / E_R )} \; = \; [\textrm{flux}]\; 
\left[
F_{2D} (x, Q^2; \alpha_R, p_{RT}) - (1-\epsilon) F_{LD} (..) \; + \; 
\ldots \right] ,
\label{cross_section}
\end{equation}
where $\epsilon$ is the virtual photon polarization parameter. 
The ``tagged'' deuteron structure functions $F_{2D}$ and $F_{LD}$ depend on
$x$ and $Q^2$ as well as on the recoil nucleon momentum, which is parametrized here by
the light-cone fraction $\alpha_R \equiv 2(E_R + p_R^z)/(E_D + p_D^z)$ and transverse
momentum $p_{RT} \equiv |\bm{p}_{RT}|$ in a frame where the momentum transfer 
$\bm{q}$ and the deuteron momentum
$\bm{p}_D$ are collinear and along the $z$--axis. Here $x \equiv Q^2/(p_D q) = 2x_{\rm B}$ 
is the rescaled Bjorken variable with limits $0 < x < 2$. 
The ellipsis in Eq.~(\ref{cross_section}) stands for azimuthal angle-- ($\phi_R$--) 
and spin--dependent terms which we left out for brevity. 
The full form of the cross section for scattering of a polarized electron
on a polarized deuteron (vector, tensor)
has been determined and will be given elsewhere \cite{Cosyn:inprep}. We emphasize that
the form of Eq.~(\ref{cross_section}) is general and does not rely on 
any assumptions regarding composite structure of the deuteron.
%
%

{\bf Light-front nuclear structure and impulse approximation.}
The theoretical description aims to express the tagged structure functions of
Eq.~(\ref{cross_section}) in terms of nuclear structure and nucleon structure functions.
High--energy processes such as DIS probe the nucleus at fixed light--front time, 
where the structure is described by the light--front wave function. The light--front 
description is unique in the sense that the off--shellness of the constituents
remains finite as the scattering energy becomes large, which permits a composite 
description of the nucleus in DIS in terms of nucleon degrees of 
freedom \cite{Frankfurt:1981mk}. It can be implemented using either non-covariant 
light--front quantization or a covariant scheme based on Feynman 
diagrams \cite{Sargsian:2001ax}. The starting point is the impulse approximation,
in which the electromagnetic current interacts with a single nucleon and the DIS final 
state evolves independently of the other nucleon (see Fig.~\ref{fig:tagged}b).
Neglecting terms $\textrm{(mass)}^2/Q^2$ it gives 
\begin{equation}
F_{2D} (x, Q^2; \alpha_R, p_{RT}) \; = \; \frac{|\Psi_{D}(\alpha_R, \bm{p}_{RT})|^2}
{2 - \alpha_R} \; F_{2N} (\widetilde x, Q^2; \; \textrm{off-shell}),
\hspace{2em} \widetilde x \; \equiv \; \frac{x}{2 - \alpha_R} .
\label{ia}
\end{equation}
Here $2 - \alpha_R$ is the LF momentum fraction of the active nucleon in the deuteron,
which is kinematically fixed by that of the spectator, $\alpha_R$. The nucleon structure
function is evaluated at $\widetilde x = x/(2 - \alpha_R)$. $\Psi_D$ is
the deuteron's $NN$ light--front wave function, normalized such that $\int d\alpha_R d^2 p_{RT}
|\Psi_{D}(\alpha_R, \bm{p}_{RT})|^2 / [\alpha_R (2 - \alpha_R)] = 1$, 
which can be obtained from the non-relativistic 
wave function for rest frame momenta $|\bm{p}_R| \ll M_N$ \cite{Frankfurt:1981mk}.
$F_{2N}$ is the structure function of the active nucleon, which is generally modified
by off-shell effects (see below). When integrated over the recoil momentum Eq.~(\ref{ia}) 
satisfies the baryon number and light-cone momentum sum rules for the deuteron.

Equation~(\ref{ia}) implies Bjorken scaling of the tagged deuteron structure function
at $Q^2 \gg 1\, \textrm{GeV}^2$ for fixed $\alpha_R$ and $p_{RT}$ (this holds even in
the presence of modified nucleon structure and final-state interactions). 
This feature can be verified experimentally 
and represents a basic test of the theoretical framework. It represents a special
case of the QCD factorization theorem for DIS with identified slow hadrons in the target 
fragmentation region \cite{Trentadue:1993ka}.

{\bf Nucleon pole and on-shell extrapolation.}
An essential feature of the impulse approximation is that it captures the leading 
singularity of the tagged structure function in the invariant momentum transfer 
between the deuteron and the recoiling nucleon, $t \equiv (p_D - p_R)^2 = 
\textrm{function}(\alpha_R, p_{RT})$. The deuteron LF wave function has a pole 
at $t = M_N^2$ corresponding to the nucleon on-shell point. (In the covariant
formulation this pole represents ``nucleon exchange'' between the deuteron and
the electromagnetic current.) In the limit $t \rightarrow M_N^2$ one therefore has
\begin{equation}
F_{2D} (x, Q^2; \alpha_R, p_{RT}) \; \sim \; \frac{R}{(t - M_N^2)^2} \; 
F_{2N} (\widetilde{x}, Q^2 ) \hspace{2em} (t \rightarrow M_N^2),
\label{pole}
\end{equation}
where $R \equiv R(\alpha_R)$ 
is a calculable residue and $F_{2N}$ is the on-shell (free) structure function 
of the active nucleon. Theoretical arguments show that final--state interactions
(see Fig.~\ref{fig:tagged}c) do not modify the leading singularity in $t$, as they 
involve an integral over the intermediate nucleon momentum \cite{Sargsian:2005rm}. 
One can thus obtain
the on-shell nucleon structure function by measuring the deuteron structure function
$F_{2D}$ over a range of $t$ at fixed $\alpha_R$ and performing the on-shell extrapolation 
to $t \rightarrow M_N^2$. The physical region is $t - M_N^2 < -\epsilon_D M_D = -0.0041 \, 
\textrm{GeV}^2$ ($\epsilon_D$ is the deuteron binding energy), so that measurements
can be done extremely close to the on-shell point. In the deuteron rest frame 
$t - M_N^2 = -2 |\bm{p}_R|^2 - \epsilon_D M_D$, and the kinematic limit corresponds to
zero recoil momentum $\bm{p}_R = 0$.

%
%
\begin{figure}
\centering
\sidecaption
\includegraphics[width=0.61\textwidth]{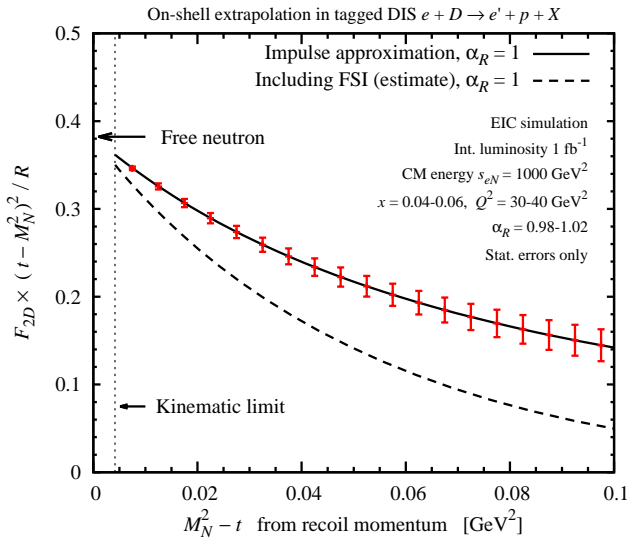}
\caption[]{Neutron structure extraction by on-shell extrapolation $t \rightarrow M_N^2$
in tagged DIS $e + D \rightarrow e' + p + X$. The plot shows the tagged deuteron structure 
function $F_{2D}$, with the pole factor $R/(t - M_N^2)^2$ removed, as a function of the 
off-shellness $t - M_N^2$, at $\alpha_R = 1$. {\it Solid line:} Impulse approximation.
{\it Dashed line:} Sum of impulse approximation and final state interactions 
(estimate) \cite{Strikman:fsi}. {\it Data:} Simulated measurement of
$t$--dependence with EIC. Parameters are shown on the plot. Error bars 
indicate expected statistical errors; for systematic errors see text.
\label{fig:onshell}}
\end{figure}
Simulations show that precise measurements of the free neutron structure function $F_{2n}$
by proton tagging and on-shell extrapolation are feasible with EIC (see Fig.~\ref{fig:onshell}).
One measures the tagged deuteron structure function $F_{2D}$ over a range of $t$ at 
fixed $\alpha_R$, removes the pole factor in Eq.~(\ref{pole}), and performs polynomial
extrapolation to $t \rightarrow 0$. Both statistical and systematic data errors can be
propagated through the extrapolation procedure. Systematic errors arise mainly from the
uncertainty in the recoil momentum determination and are expected to result in a correlated
$F_{2n}$ uncertainty of $\lesssim 3\%$ with the JLab EIC beam and detector 
design \cite{LD1506}.

The on-shell extrapolation can also be used to extract neutron spin
structure functions from polarized tagged DIS on the deuteron.
It is convenient to measure the double spin asymmetry of the tagged cross section
as a function of the recoil momentum,
\begin{eqnarray}
A_\parallel (x, Q^2; \alpha_R, p_{RT}) 
\;\; = \;\; 
\frac{\sigma (+\frac{1}{2}, +1) - \sigma (-\frac{1}{2}, +1)}
{\sigma (+\frac{1}{2}, +1) + \sigma (-\frac{1}{2}, +1)} ,
\end{eqnarray}
and perform the extrapolation to $t \rightarrow M_N^2$ for the asymmetry rather 
than the cross sections. [Here 
$\sigma (\lambda_e, \lambda_D)$ denotes the cross section for electron helicity 
$\lambda_e = \pm 1/2$ and deuteron helicity $\lambda_{D} = \pm 1$,
i.e., longitudinal polarization along the respective beam directions.]
The asymmetry exhibits only very weak $t$ dependence.
An interesting feature is that the D--wave in the deuteron 
wave function practically drops out in the
on-shell extrapolation, since its wave function is proportional to the squared 
rest frame nucleon momentum $|\bm{p}_R|^2$ and the on-shell point corresponds 
to extremely small unphysical momenta $|\bm{p}_R|^2 = -\epsilon_D M_D$.
This means that at the on-shell point the neutron is effectively 100\% polarized 
in the deuteron spin direction. Furthermore, kinematic factors and many systematic 
errors cancel in the cross section ratio. Simulations show that precise measurements 
of neutron spin structure with on-shell extrapolation are possible over most of 
the ($x, Q^2$) range covered by EIC with an integrated luminosity
$\sim 100 \, \textrm{fb}^{-1}$ \cite{Cosyn:2014zfa}.

{\bf Nuclear modifications and final--state interactions (FSI).}
Away from the nucleon pole $t = M_N^2$ the tagged deuteron structure function is 
modified both by possible off--shell dependence of the nucleon structure functions 
and by FSI. The off-shell modifications 
can be accommodated within the impulse approximation, by allowing the effective 
nucleon structure function to depend on $t - M_N^2$ in a certain range of 
virtualities (``virtual nucleon model''). Specific dynamical
models for the off-shell dependence have been proposed.

FSI arise from the interactions of the spectator with the DIS
final state produced by the active nucleon (see Fig.~\ref{fig:tagged}c).
While they do not change the total (untagged) DIS cross section on the deuteron, 
such interactions can distort the recoil momentum spectrum and change the outgoing 
particle flux, and must therefore be accounted for in the analysis of tagged DIS.
FSI can be estimated theoretically by modeling the composition
of the nucleon DIS final state and its rescattering from the spectator.
Theoretical arguments suggest that at $x \lesssim 0.1$ the dominant FSI arise
from ``slow'' hadrons in the DIS final state, with rest frame momenta $\sim$ 1--2 GeV/$c$,
as these hadrons are expected to form close to the interaction point and can
rescatter from the spectator with hadronic cross sections \cite{Strikman:fsi}. 
``Fast'' hadrons with momenta $\gg 1$ GeV/$c$ form outside the nucleus
and interact marginally with the spectators, as evidenced by the absence of
absorption of such hadrons in DIS on larger nuclei \cite{Ashman:1991cx}. Estimates 
of the FSI effects in tagged DIS can be made in a schematic model, using empirical 
DIS hadron distributions and rescattering cross sections \cite{Strikman:fsi}. 
Preliminary results show noticeable modifications of the $t$--dependence of the tagged 
structure function away from the pole (see Fig.~\ref{fig:onshell}). The modifications
vanish in the limit $t \rightarrow M_N^2$, in accordance with the findings of
Ref.~\cite{Sargsian:2005rm}, and thus do not prevent the extraction of free neutron
structure through on-shell extrapolation; their effect on the extraction errors
remains to be investigated.\footnote{The FSI effect on the $t$--distribution
depends on the recoil light-cone fraction $\alpha_R$ \cite{Strikman:fsi}. FSI is maximal for 
the situation shown in Fig.~\ref{fig:onshell}, $\alpha_R = 1$, which corresponds 
to an angle of $\sim 90^\circ$ between the recoil momentum $\bm{p}_R$ and the $\bm{q}$--vector
in the rest frame (sideways recoil). FSI becomes smaller for $\alpha_R < 1$ or $>1$,
which correspond to angles $<90^\circ$ or $>90^\circ$ (forward or backward recoil).
However, for $\alpha_R \neq 1$ the kinematic limit in $t - M_N^2$ moves further
away from zero, which is not favorable for on-shell extrapolation.}

An interesting question is how the effects of modified nucleon structure and FSI
in tagged DIS could be separated experimentally in future measurements at EIC. 
Tagged DIS measurements at $\widetilde x \sim 0.2$, where nuclear modifications of the 
inclusive structure functions are negligible \cite{Malace:2014uea}, could isolate the 
effects of FSI in the tagged DIS cross section. Direct evidence for the dominance of 
FSI in this region could come from observation of their distinctive angular dependence 
at larger rest frame recoil momenta $p_R \gtrsim$ 200 MeV/$c$ \cite{Strikman:fsi}. 
Tagged DIS measurements at $\widetilde x > 0.3$, where substantial nuclear 
modifications are seen in the inclusive data (EMC effect) \cite{Malace:2014uea}, 
could then reveal the combined effects of modified nucleon structure and FSI.
Further information on nuclear FSI will come from measurements of nuclear modification
of hadron spectra in DIS on heavier nuclei. The experimental capabilities of EIC (kinematic
range, recoil momentum coverage and resolution, variety of nuclear beams) would enable a
program of comparative measurements designed to separate the two effects.

\begin{figure}
\centering
\includegraphics[width=0.9\textwidth]{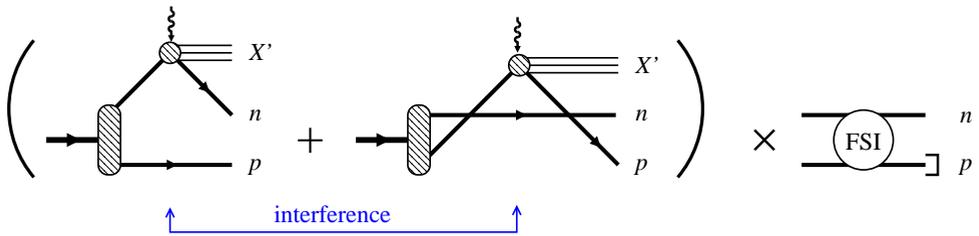}
\caption[]{Diffraction in tagged DIS on the deuteron $e + D \rightarrow e' + p + X$.
Diffractive scattering on the $p$ and $n$ enables interference of the respective amplitudes 
(shadowing), as well as strong FSI within the outgoing $pn$ system.
\label{fig:diffract}}
\end{figure}

{\bf Diffraction in tagged DIS.}
In tagged DIS at $x \lesssim 0.01$ a new mechanism of FSI arises due to diffractive
scattering on the active nucleon, in which the nucleon is left intact and appears
in the DIS final state with a rest--frame momentum 
$\sim$ few 100 MeV/$c$ (see Fig.~\ref{fig:diffract}).
In DIS on the deuteron such diffractive scattering can happen on the proton as well
as on the neutron, with quantum-mechanical interference of the two amplitudes in the
cross section. In inclusive DIS $e + D \rightarrow e'+ X$ this effect gives rise to
nuclear shadowing and has been the subject of extensive theoretical 
studies \cite{Frankfurt:2011cs}. In tagged DIS $e + D \rightarrow e'+ N + X$
the presence of the slow diffractive nucleon in the final state enables strong FSI
between that nucleon and the spectator, resulting in considerable modification of the
recoil momentum distribution compared to the impulse approximation. The distortion
is particularly strong in the spin--1, isospin--0 partial wave of the $pn$ system,
where the outgoing $pn$ scattering state must be orthogonal to the deuteron bound 
state in that channel (a similar effect happens in low--energy deuteron 
breakup reactions). Theoretical work aims to calculate this effect using
methods analogous to those used for shadowing in inclusive DIS \cite{guzey:inprogress}.

Experiments at EIC could perform detailed tests of the predicted distortion
effects using single--nucleon tagging $e + D \rightarrow e'+ p + X$, or 
double--nucleon tagging $e + D \rightarrow e'+ p + n + X'$, which completely fixes the
kinematics of the produced $pn$ system. In this way they could elucidate the general
mechanism of nuclear shadowing, which is essential for analyzing inclusive scattering
on the deuteron (including polarization) and quantifying the approach to saturation
at small $x$.
%
%

In summary, electron--deuteron DIS with spectator tagging at EIC would enable 
next--generation measurements of neutron spin structure, nuclear modifications
of partonic structure, and small--$x$ shadowing, with maximum control of the nuclear
configurations participating in the high--energy process. Theoretical methods based
on light--front nuclear structure and analytic properties provide a transparent and efficient 
description of such measurements, consistent with that of inclusive nuclear DIS.
Work is in progress to adapt these methods to collider kinematics and polarized beams.

Notice: Authored by Jefferson Science Associates, LLC
under U.S.~DOE Contract No.~DE-AC05-06OR23177.
The U.S.~Government retains a non-exclusive, paid-up,
irrevocable, world-wide license to publish or reproduce
this manuscript for U.S.~Government purposes.
\end{document}